\documentclass[aps,pra,preprint,groupedaddress,showpacs,%
tightenlines,%
nofootinbib,floatfix]{revtex4-1}

\usepackage{upgreek}  
\usepackage{graphicx}
\usepackage{amsmath,amssymb,amsfonts,bm}
\usepackage{array,dcolumn}
\usepackage{hyperref}
\usepackage{ulem} 

\renewcommand{\vec}[1]{\bm{\mathrm{#1}}} 

\begin{document}

\title{An influence of the spectator-nuclear motion on nonresonant
  formation of the muonic hydrogen molecules}

\author{Andrzej Adamczak}
\email{andrzej.adamczak@ifj.edu.pl}
\affiliation{Institute of Nuclear Physics Polish Academy of Sciences,
  Radzikowskiego 152, PL-31342~Krak\'ow, Poland}

\author{Mark P.\ Faifman}
\email{ faifmark@gmail.com}

\affiliation{National Research Centre Kurchatov Institute, pl.~Akademika
  Kurchatova~1, 123182 Moscow, Russia}

\date{\today}

\begin{abstract}
  A model for description of nonresonant formation of the muonic
  molecules in collisions of the muonic hydrogen atoms with the
  hydrogenic molecules has been developed with taking into account the
  internal motion of all nuclei. It has been shown that such a motion
  leads to a significant smearing of the calculated energy-dependent
  formation rates at low collision energies. In particular, this effect
  is strong in the $dd\mu$ and $dt\mu$ formation. An appreciable
  isotopic effect in the case of nonresonant $dd\mu$ formation in $d\mu$
  collisions with the molecules D$_2$ and HD has been found. All these
  effects are of importance for many experimental researches in
  low-energy muon physics.
\end{abstract}


\maketitle

\section{Introduction}
\label{intro}

The formation of muonic molecules is one of crucial links in the chain
of physical processes caused by the negative $\mu^{-}$ muons in
a~hydrogen isotope mixtures (see reviews~\cite{zg,pon90,bal11} and
references therein). The interpretation and analysis of the data
obtained in various experiments with low-energy muons require a
knowledge of the energy-dependent formation rates of various muonic
hydrogen molecules. For example, in the studies of muon-catalyzed $pt$
and $tt$ fusion~\cite{bogd12,bogd15}, in the PSI
measurements~\cite{mucap,musun} of the muon capture in the
hydrogen-isotope nuclei, as well as in the planned determination of the
Zemach radius of proton by the FAMU collaboration~\cite{adam12} at RAL
and by the CREMA collaboration~\cite{crema2017} at PSI.

The muonic molecule being the three-body system (ion, in reality)
consists of the two hydrogen isotope nuclei and the muon. Such
$\mu$-molecular systems are formed in collisions of the muonic
$a\mu$-atoms ($a=p$, $d$, or~$t$) with the hydrogen-isotope (hydrogenic)
molecules~$BX$ ($B$, $X=$~H, D, or~T). The quantum states of formed
muonic molecules are defined by the different rotational~($J$) and
vibrational~($\upsilon$) quantum numbers
(see~Table~\ref{tab:1})\footnote{The table is compiled from the data of
  Ref.~\cite{vin}, whereas the binding energies of the state ($J=1$,
  $\upsilon=1$) are taken from Ref.~\cite{korob} (see also
  Ref.~\cite{frol}).}.
\begin{table}[htb]
  \begin{center}
    \caption{The binding energies $|\varepsilon_{J\upsilon}|$ (in eV) of
      the muonic molecules $ab\mu$ in the states ($J\upsilon$).}
    \label{tab:1}
    \begin{ruledtabular}
      \newcolumntype{.}{D{.}{.}{3.3}}
      \begin{tabular}{c. . . . . .}
      \multicolumn{1}{c}{State~~}&\multicolumn{6}{c}{Molecule}\\[3pt]
      \multicolumn{1}{c}{($Jv$)}&
      \multicolumn{1}{c}{$pp\mu$}&
      \multicolumn{1}{c}{$pd\mu$}&
      \multicolumn{1}{c}{$pt\mu$}&
      \multicolumn{1}{c}{$dd\mu$}&
      \multicolumn{1}{c}{$dt\mu$}&
      \multicolumn{1}{c}{$tt\mu$}\\[3pt]
      \hline\noalign{\vskip2pt}
(11) & \multicolumn{1}{c}{---}& \multicolumn{1}{c}{---}& 
       \multicolumn{1}{c}{---}& 1.965   & 0.631   & 45.206  \\
(30) & \multicolumn{1}{c}{---}& \multicolumn{1}{c}{---}& 
       \multicolumn{1}{c}{---}& \multicolumn{1}{c}{---}& 
       \multicolumn{1}{c}{---}& 48.838  \\
(01) & \multicolumn{1}{c}{---}& \multicolumn{1}{c}{---}& 
       \multicolumn{1}{c}{---}& 35.844  & 34.834  & 83.771  \\
(20) & \multicolumn{1}{c}{---}& \multicolumn{1}{c}{---}& 
       \multicolumn{1}{c}{---}& 86.494  & 102.649 & 172.702 \\
(10) & 107.266  & 97.498   & 99.126  & 226.682 & 232.471 & 289.142 \\
(00) & 253.152  & 221.549  & 213.840 & 325.074 & 319.140 & 362.910 \\
      \end{tabular}
    \end{ruledtabular}
  \end{center}
\end{table}%

The loosely bound ($J=1$, $\upsilon=1$) states of the $dd\mu$ and
$dt\mu$ molecules refer to the formation, as a rule, by the resonance
mechanism~\cite{ves} in the reaction of following type
\begin{equation}
  \label{eq1}
  t\mu + \mathrm{D}_2\rightarrow[(dt\mu)_{11}\, dee]_{K\nu}^{*} \,,
\end{equation}
where the released energy of about $|\varepsilon_{11}|$ is transferred
to the excitation of rotational-vibrational states ($K\nu$) of the
molecular complex $[(dt\mu)_{11}dee]$. The rates~$\lambda_{dt\mu}$ and
$\lambda_{dd\mu}$ of such resonance reactions depend on the target
temperature~$T$. For the room temperature $T=300$~K, these rates are on
the order of $10^8$~s$^{-1}$~\cite{pon90} and
$10^6$~s$^{-1}$~\cite{bal11,men87}, respectively.

In any other state ($J\upsilon$), the muonic molecular ions $ab\mu$ are
formed via the nonresonant process~\cite{zg,cohe60,pono76}
\begin{equation}
  \label{eq2}
  a\mu + BX \to [(ab\mu)_{J\upsilon}\, xe]^{+}+e^{-} \,,
\end{equation}
with conversion of the released energy into electron ionization of the
$BX$ molecule. The rates of transitions~(\ref{eq2}) to all
($J\upsilon$)-states of $ab\mu$ have been calculated in
Ref.~\cite{mpf89}, except that for the ($J=1$, $\upsilon=1$) states of
the molecules $dd\mu$ and $dt\mu$. Later on, the nonresonant formation
in such a loosely bound state was also considered for collision energies
higher than the electron ionization potential of the $BX$
molecule~\cite{af12}. There was shown that the corresponding rates of
reactions~(\ref{eq2}) could be significant. The rates of such
nonresonant formation reach the magnitudes up to $10^7$~s$^{-1}$,
depending on energy of $a\mu$ collision with a~$BX$ molecule.

A comparison of the measured~\cite{bal11,kn12,mucap15} and
calculated~\cite{mpf89,af12} rates of nonresonant transitions in
reactions~(\ref{eq2}) demonstrated a very good agreement for different
muonic molecules, although some differences between the theory and
experiments were observed in the case of low-temperature H/D targets. In
order to improve the calculating scheme~\cite{mpf89}, in which the
distance between the center of mass (CM) of molecule $ab\mu$ and the
spectator nucleus~$x$ of the molecule $BX$ was kept constant, the
internal motion of both the nuclei $b$ and $x$ within the $BX$ molecule
is taken into account in the present work.

In Sec.~\ref{sec:trans_cms_lab}, a transformation of the nonresonant
formation rates calculated~\cite{mpf89} in the center of mass of the
$a\mu+b$ system to the laboratory system $a\mu+BX$ is carried out with
taking into account the internal motion of nuclei $b$ and~$x$ within the
molecule~$BC$. A harmonic model of the molecular vibrations is used in
Sec.~\ref{sec:nuc_motion} in order quantitatively to describe the
internal nuclear motion within the hydrogenic molecules. The results and
discussion are included in Sec.~\ref{sec:discussion}.

\section{Transformation of the nonresonant formation rates
 between the center-of-mass and laboratory systems}
\label{sec:trans_cms_lab}

\begin{figure}[htb]
  \centering
  \includegraphics[width=5cm]{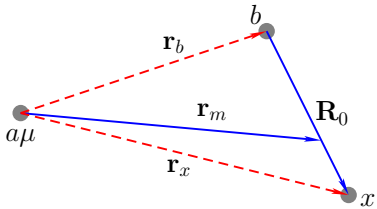}
  \caption{(Color online) Relative coordinates for the description of
    nonresonant formation of the $ab\mu$ molecule in $a\mu$ collision
    with molecule~$BX$. The electrons are not displayed here.
    Vector~$\vec{r}_m$ denotes the position of the $BX$ molecule center
    of mass reckoned from the $a\mu$ atom center of mass.}
  \label{sys_mol}
\end{figure}

The rates~$\lambda$ of nonresonant formation of $ab\mu$ molecules were
calculated in Ref.~\cite{mpf89} assuming that the nucleus~$x$ in the
molecule $BX$ is a distant spectator, which is located at a~fixed
position~$\vec{R}_0$ with respect to the nucleus~$b$ (see
Fig.~\ref{sys_mol}).
These rates, which were calculated in the center of mass of the $a\mu+b$
system (\textit{nuclear} CMS), are functions of collision energy
$\varepsilon$ in this system: $\lambda=\lambda(\varepsilon)$.
A~dependence of $\varepsilon$ on the momentum $\vec{k}$ of relative
motion of the $a\mu$ atom and the nucleus~$b$ is expressed as follows
\begin{equation}
  \label{eq3}
  \varepsilon = \frac{\vec{k}^2}{2\mu} \,,  \qquad
  \vec{k} = \mu \frac{\mathrm{d}\vec{r}_b}{\mathrm{d}t} \,,
\end{equation}
where the reduced mass $\mu$ of the $a\mu+b$ system is given by the
relations
\begin{equation}
  \label{eq4}
  \mu^{-1} = M_{a\mu}^{-1}+M_b^{-1}, \qquad  M_{a\mu}=M_a+M_\mu \,,
\end{equation}
in which $M_\mu$ is the muon mass and the masses of nuclei $a$ and $b$
are denoted by $M_a$ and $M_b$, respectively.

For the analysis of the experimental data, it is often more convenient
to use the formation rates as functions of kinetic energy~$E$ of the
$a\mu$ atoms in the laboratory frame~(LAB). Such LAB rates were
estimated in Refs.~\cite{af12,af09} using the rates
$\lambda(\varepsilon)$, which were calculated in the nuclear CMS, and
the transformation of the following form~\cite{fms89,pet94}:
\begin{equation}
  \lambda(E,T) = \int_{0}^{\infty} \lambda(\tilde{\varepsilon}_Q)
  \, G(\vec{p},\tilde{\varepsilon}_Q) \,
  \mathrm{d}\tilde{\varepsilon}_Q \,,
  \label{eq5}
\end{equation}
with the ``distribution function''~$G(\vec{p},\varepsilon_Q)$ defined
below
\begin{equation}
\label{eq6}
\begin{split}
G(\vec{p},\varepsilon_Q) & =
\sqrt{\frac{M_{bx}}{2\pi T}}\frac{M_{a\mu}}{\mu_t p}
\left\{\exp\left[-\displaystyle{\frac{M_{bx}}{2T}\left(\frac{Q}{\mu_t}
-\frac{p}{M_{a\mu}}\right)^{\!2}}\right]\right. \\ &
-\exp\left.\left[-\displaystyle{\frac{M_{bx}}{2T}\left(\frac{Q}{\mu_t}
+\frac{p}{M_{a\mu}}\right)^{\!2}}\right]\right\}.
\end{split}
\end{equation}
The total mass $M_t$ and the reduced mass $\mu_{t}$ of the $a\mu+BX$
system (henceforth called the \textit{molecular} system) are given by
the relations
\begin{equation}
  \label{eq7}
  \begin{split}
    M_t & = M_{a\mu}+M_{bx} \,, \quad
    \mu_t^{-1} = M_{a\mu}^{-1}+M_{bx}^{-1} \,, \\
    M_{bx} & = M_b+M_x \,,
\end{split}
\end{equation}
where $M_x$ denotes the mass of nucleus~$x$.
In~Eqs.~(\ref{eq5})--(\ref{eq6}), the momentum $\vec{p}$ of $a\mu$ atom
in~LAB and the momentum $\vec{Q}$ of relative motion of the $a\mu$ atom
and molecule $BX$ are related to the corresponding kinetic energies $E$
and $\varepsilon_Q$ by the following expressions:
\begin{equation}
  E = \vec{p}^2/(2M_{a\mu})
  \label{eq8}
\end{equation}
and
\begin{equation}
  \varepsilon_Q = \frac{\vec{Q}^2}{2\mu_{t}} \,, \qquad
  \vec{Q} = \mu_{t}\frac{\mathrm{d}\vec{r}_m}{\mathrm{d}t} \,.
  \label{eq9}
\end{equation}

The employment of Eqs.~(\ref{eq5})--(\ref{eq9}) for calculating the
formation rates was based on the assumption that the target molecules
$BX$ are point-like objects and their kinetic-energy distribution in
a~target at temperature~$T$ has the Maxwellian shape. The dependencies
\begin{equation}
  \varepsilon_Q=\frac{\mu_t}{\mu}\varepsilon \,,
  \qquad \vec{Q}=\frac{\mu_t}{\mu}\, \vec{k}
  \label{eq10}
\end{equation}
between the relative energies $\varepsilon_Q$ and $\varepsilon$, as well
as between the corresponding momenta $\vec{Q}$ and $\vec{k}$, are
inferred from Eqs.~(\ref{eq3}) and~(\ref{eq9}) using the approximation
$\vec{r}_m\approx\vec{r}_b$.

The simplified procedure of the formation-rate transformation between
the \textit{nuclear} (or \textit{molecular}) CMS and LAB, which is
briefly described above, denotes in fact that the internal motion of the
nuclei in a~target molecule $BX$ is neglected and that merely the
correct masses~$M_{bx}$ and~$\mu_t$ of the system $a\mu+BX$, which are
defined in Eq.~(\ref{eq7}), are substituted in
Eqs.~(\ref{eq5})--(\ref{eq8}) instead of the masses $M_b$ and~$\mu$ of
the system $a\mu+b$ [from Eq~(\ref{eq4})].

Since the hydrogenic molecules are very light, the kinetic
energy~$\varepsilon_{q}$
\begin{equation}
  \varepsilon_{q} = \frac{\vec{q}^2}{2\mu_{bx}} \,, \quad
  \vec{q} = \mu_{bx}\frac{\mathrm{d}\vec{R}_0}{\mathrm{d}t}
  \,,\quad
  \mu_{bx}^{-1} = M_b^{-1}+M_x^{-1} \,,
  \label{eq11}
\end{equation}
which corresponds to the relative momentum $\vec{q}$, is quite large.
Even in the ground state of the molecule $BX$, this energy is on the
order of its vibrational quantum $\omega_0\approx{}0.3$--0.5~eV
\cite{ll89} due to the zero-point vibrations. Therefore, the previous
above-mentioned scheme of $\lambda(\varepsilon)$ transformation between
CMS and LAB is not valid in general. In order to improve this scheme, it
is indispensable to consider the $a\mu$ atom collision with a~real $BX$
molecule of finite dimensions, instead of a point-like molecule $BX$
that is located at the position of a free nucleus~$b$.

The following relations between different vectors of the
$a\mu+BX$ system are inferred from~Fig.~\ref{sys_mol}:
\begin{align}
  &\vec{R}_0 =\vec{R}_x-\vec{R}_b =\vec{r}_x-\vec{r}_b\,,&
  & & \nonumber \\[5pt]
  &\vec{R}_m =\frac{M_b \vec{R}_b+M_x \vec{R}_x}{M_{bx}}\,,&
  &\vec{r}_m =\frac{M_b \vec{r}_b+M_x \vec{r}_x}{M_{bx}}\,,&
  \nonumber \\[5pt]
  &\vec{R}_b =\vec{R}_m-\frac{M_x}{M_{bx}} \vec{R}_0\,,&
  &\vec{r}_b =\vec{r}_m-\frac{M_x}{M_{bx}}\vec{R}_0\,,&
  \label{eq12} \\
  &\vec{R}_x =\vec{R}_m+\frac{M_b}{M_{bx}}\vec{R}_0\,,&
  &\vec{r}_x =\vec{r}_m+\frac{M_b}{M_{bx}}\vec{R}_0\,, \nonumber
\end{align}
in which vectors $\vec{R}_b$, $\vec{R}_x$, and $\vec{R}_m$ denote the
positions of the nuclei $b$ and $x$, and their center of mass in~LAB,
respectively. Vectors $\vec{r}_b$, $\vec{r}_x$ and $\vec{r}_m$
correspond to the relative positions of the particles.

The momentum $\vec{p}$ of $a\mu$ atom and the momentum $\vec{p}_b$ of
nucleus~$b$ in LAB are connected with the relative momentum $\vec{k}$
[see Eq.~(\ref{eq3})] by the following formulas~\cite{ll89}:
\begin{align}
  \label{eq13}
  &\vec{p} =-\vec{k}+\frac{M_{a\mu}}{M}\vec{\mathcal{P}} \,, \\[5pt]
  \label{eq14}
  &\vec{p}_b =\vec{k}+\frac{M_b}{M}\vec{\mathcal{P}} \,,
\end{align}
where $M=M_a+M_b+M_{\mu}$ and $\vec{\mathcal{P}}=\vec{p}+\vec{p}_b$ are
the total mass and momentum of the 3-body system ($ab\mu$),
respectively. The inverse relation for the momentum $\vec{k}$ is evident
from~Eqs.~(\ref{eq13})--(\ref{eq14}):
\begin{equation}
\vec{k}=(M_{a\mu}{\vec{p}_b}-M_b{\vec{p}})/M \,.
  \label{eq15}
\end{equation}
Analogously to Eqs.~(\ref{eq13})--(\ref{eq14}), the following relations
\begin{align}
  \label{eq16}
  &\vec{p}_b = -\vec{q}+\frac{M_b}{M_{bx}}\vec{P}_m \,, \\[5pt]
  \label{eq17}
  &\vec{p}_x = \vec{q}+\frac{M_x}{M_{bx}}\vec{P}_m \,, \\[5pt]
  \label{eq18}
  &\vec{P}_m=\vec{p}_x+\vec{p}_b \,,
\end{align}
hold for the LAB momenta $\vec{p}_b$ and $\vec{p}_x$ of the respective
nuclei $b$ and~$x$, and the vector $\vec{P}_m$ of the total momentum of
molecule $BX$. The relative momentum $\vec{q}$ of the nuclei $b$ and $x$
[see Eq.~(\ref{eq11}) and Fig.~\ref{sys_mol}] is given by the expression
\begin{equation}
  \vec{q} = (M_b{\vec{p}_x}-M_x{\vec{p}_b})/M_{bx} \,,
  \label{eq19}
\end{equation}
which is similar to Eq.~(\ref{eq15}). By virtue of Eq.~(\ref{eq16}),
the momentum~(\ref{eq15}) takes the following form:
\begin{equation}
  \vec{k} = \frac{1}{M} \left( \frac{M_{a\mu}M_b}{M_{bx}}\,\vec{P}_m
    - M_{a\mu}{\vec{q}} - M_b{\vec{p}} \right) .
  \label{eq20}
\end{equation}
From Fig.~\ref{sys_mol}, one can infer the relation
\begin{equation}
  \vec{r}_b = \vec{r}_m - \beta_b\vec{R}_0  \,,
  \label{eq21}
\end{equation}
where $\beta_b\vec{R}_0$ denotes the position of the molecule $BX$
center of mass with respect to nucleus~$b$:
\begin{equation}
  \label{eq22}
  \beta_b = \frac{M_x}{M_b+M_x} = \frac{\mu_{bx}}{M_b} \,.
\end{equation}
The equation~(\ref{eq21}), together with the definitions~(\ref{eq3}),
(\ref{eq9}) and~(\ref{eq11}), leads to the following relation
\begin{equation}
  \vec{k} = \mu \left(\frac{\vec{Q}}{\mu_t}
    - \beta_b\frac{\vec{q}}{\mu_{bx}} \right)
  \label{eq23}
\end{equation}
between the momenta $\vec{k}$, $\vec{q}$ and~$\vec{Q}$. Analogously
to~Eq.~(\ref{eq3}), the relative kinetic energy~$\varepsilon$ can be
expressed as
\begin{equation}
  \varepsilon\equiv\,\varepsilon(\vec{Q},\vec{q}) = \frac{\mu}{2}
  \left(\frac{\vec{Q}}{\mu_t}-\beta_b \frac{\vec{q}}{\mu_{bx}}\right)^2.
  \label{eq24}
\end{equation}

A relation between the formation rate $\lambda(\varepsilon)$ in the
center of mass of the $a\mu+b$ system and the respective rate
$\lambda(E)$ in the LAB frame of the $a\mu+BX$ system, which takes into
account the internal motion of nucleus~$b$ within~$BX$, can be written
down in the form
\begin{equation}
\lambda(E,T)\equiv\lambda(\vec{p},T)=\int{\lambda(\varepsilon) \,
\mathrm{dW}(\vec{p}_b,\vec{p}_x)\, \mathrm{d}\mathrm{F}_T(\vec{P})} \,,
  \label{eq25}
\end{equation}
where $\mathrm{F}_T(\vec{P})\equiv\mathrm{F}(\vec{P},T)$ is the
Maxwellian distribution of the molecules~$BX$ over their
momenta~$\vec{P}$ in~LAB
\begin{equation}
  \begin{split}
    \mathrm{d}\mathrm{F}_T(\vec{P}) & = \mathrm{F}(\vec{P},T)\,
    \mathrm{d}^3P \\
    & = (2\pi M_{bx}T)^{-3/2}\exp{(-\vec{P}^{2}/2M_{bx}T)}\,
    \mathrm{d}^3P \,,
  \end{split}
  \label{eq26}
\end{equation}
and $\mathrm{W}(\vec{p}_b,\vec{p}_x)$ is the distribution of nuclei $b$
and~$x$
\begin{equation}
  \mathrm{d}\mathrm{W}(\vec{p}_b,\vec{p}_x) =
  |\Psi(\vec{p}_b,\vec{p}_x)|^2 \frac{\mathrm{d}^3p_b}{(2\pi)^3}\,
  \frac{\mathrm{d}^3p_x}{(2\pi)^3}
  \label{eq27}
\end{equation}
with respect to the momenta $\vec{p}_b$ and~$\vec{p}_x$ in the LAB
frame. Function $\Psi(\vec{p}_b,\vec{p}_x)$ denotes here the wave
function of molecule $BX$ in the momentum representation, which is
equivalent to the real-space function
\begin{equation}
  \Psi(\vec{r}_b,\vec{r}_x) =
  \exp(i\vec{P}\cdot\vec{R}_m)\Phi(\vec{R}_0) \,,
  \label{eq28}
\end{equation}
with both the total momentum $\vec{P}$ and the center-of-mass
position~$\vec{R}_m$ of $BX$ given in~LAB. In the momentum
representation, the function (\ref{eq28}) has the following form:
\begin{equation}
  \begin{split}
    \Psi(\vec{p}_b,\vec{p}_x) = \int & \Psi(\vec{r}_b,\vec{r}_x)
    \exp(-i\vec{p}_b\cdot\vec{r}_b) \\
    & \exp(-i\vec{p}_x\cdot\vec{r}_x)
    \, \mathrm{d}^3r_b\, \mathrm{d}^3r_x \,.
\end{split}
\label{eq29}
\end{equation}
Upon employing the expression
\begin{equation}
  \vec{p}_b\cdot\vec{r}_b+\vec{p}_x\cdot\vec{r}_x =
  \vec{q}\cdot\vec{R}_0+\vec{P}_m\cdot\vec{R}_m \,,
  \label{eq30}
\end{equation}
which follows from Eqs.~(\ref{eq12})--(\ref{eq14}), and changing the
variables
$\mathrm{d}^3r_b\,\mathrm{d}^3r_x\rightarrow\mathrm{d}^3R_m\,\mathrm{d}^3R_0$
in~Eq.~(\ref{eq29}), we obtain
\begin{equation}
  \begin{split}
    \Psi(\vec{p}_b,\vec{p}_x) = \int & \Phi(\vec{R}_0)
    \exp(-i\vec{q}\cdot\vec{R}_0) \\
    & \exp[-i(\vec{P}_m-\vec{P})\cdot\vec{R}_m]
    \,\mathrm{d}^3R_0\,\mathrm{d}^3R_m \,.
  \end{split}
  \label{eq31}
\end{equation}
Then, using the following representation of the Dirac delta function
\begin{equation}
  \int \exp[-i(\vec{P}_m-\vec{P})\cdot\vec{R}_m]\, \mathrm{d}^3R_m
  = (2\pi)^3\delta(\vec{P}_m-\vec{P}) \,
  \label{eq32}
\end{equation}
and the definition
\begin{equation}
  \psi(\vec{q})\equiv\int{\Phi(\vec{R}_0)\exp(-i\vec{q}\cdot\vec{R}_0)
    \, \mathrm{d}^3R_0} \,,
  \label{eq33}
\end{equation}
equation~(\ref{eq31}) leads to the relation
\begin{equation}
  |\Psi(\vec{p}_b,\vec{p}_x)|^2=[(2\pi)^3\delta(\vec{P}_m-\vec{P})]^2
  |\psi(\vec{q})|^2 \,.
  \label{eq34}
\end{equation}
This relation can be simplified on employing the following
formula~\cite{r82}:
\begin{equation}
  \delta(x-a)\delta(x-b)=V\delta(x-a) \,,
  \label{eq35}
\end{equation}
in which the arbitrary constant corresponds to the normalization of
volume: $V=1$. Then, the expression (\ref{eq34}) can be written as
\begin{equation}
  |\Psi(\vec{p}_b,\vec{p}_x)|^2=(2\pi)^3\delta(\vec{P}_m-\vec{P})
  |\psi(\vec{q})|^2 .
  \label{eq36}
\end{equation}
The formation rate~(\ref{eq25}) in LAB now takes the form
\begin{equation}
  \lambda(\vec{p},T)=\int{\lambda(\varepsilon)\, \mathrm{F}(\vec{P},T)\,
    \mathrm{d}^3P\, |\psi(\vec{q})|^2 \,
    \frac{\mathrm{d}^3q}{(2\pi)^3}} \,,
  \label{eq37}
\end{equation}
by virtue of Eqs.~(\ref{eq26}), (\ref{eq27}), (\ref{eq36}), and the
subsequent change $\mathrm{d}^3p_b\,\mathrm{d}^3p_x\rightarrow %
\mathrm{d}^3P_m\,\mathrm{d}^3q$,
of the integration variables and the integration over $\mathrm{d}^3P_m$.

Using Eqs.~(\ref{eq18}) and (\ref{eq19}) the collision energy
$\varepsilon$ defined in Eq.~(\ref{eq3}) can be expressed in terms of
the LAB momenta
\begin{equation}
  \varepsilon=\frac{1}{2\mu {M}^2}
  \left( M_b{\vec{p}}+M_{a\mu}{\vec{q}}-\frac{M_{a\mu}M_b}{M_{bx}}
    \vec{P}\right)^2 .
  \label{eq38}
\end{equation}
Now, the rate~(\ref{eq37}) can be written down as follows~\cite{fms89}:
\begin{equation}
  \lambda(E,T)\equiv \lambda(\vec{p},T)=\int \mathrm{F}(\vec{P},T)\,
  \Lambda(\varepsilon_Q)\, \mathrm{d}^3P\,,
  \label{eq39}
\end{equation}
where
\begin{equation}
  \begin{split}
    \Lambda(\varepsilon_Q) & \equiv
    \int |\psi(\vec{q})|^2\, \frac{\mathrm{d}^3q}{(2\pi)^3}
    \int_{0}^{\infty} \delta(\varepsilon-\tilde{\varepsilon}) \,
    \lambda(\tilde{\varepsilon}) \, \mathrm{d}\tilde{\varepsilon} \\
    & = \int_{0}^{\infty} \lambda(\tilde{\varepsilon})\,
    g(\varepsilon_Q,\tilde{\varepsilon})\,
    \mathrm{d}\tilde{\varepsilon}\,,
  \end{split}
  \label{eq40}
\end{equation}
and the notation
\begin{equation}
  g(\varepsilon_Q,\tilde{\varepsilon})\equiv\int
  |\psi(\vec{q})|^2 \delta(\varepsilon-\tilde{\varepsilon})\,
  \frac{\mathrm{d}^3q}{(2\pi)^3}
  \label{eq41}
\end{equation}
is introduced. The characteristic quantity $\Lambda(\varepsilon_Q)$ in
Eq.~(\ref{eq39}) can be considered as the ``formation rate'' in the
\textit{molecular} CMS, which has been obtained upon the rate
transformation from the \textit{nuclear} CMS and then averaged over the
momenta $\vec{q}$ of the nuclei with the distribution
function~$g(\varepsilon_Q,\tilde{\varepsilon})$, in accordance with
Eqs.~(\ref{eq38}) and~(\ref{eq11}). Finally, Eq.~(\ref{eq39}) for the
formation rate in~LAB is reduced to the form similar to that of
formula~(\ref{eq5}):
\begin{equation}
  \begin{split}
    \lambda(\vec{p},T) & =\int \mathrm{F}(\vec{P},T)\, \mathrm{d}^3P
    \int_{0}^{\infty} \delta(\varepsilon_Q-\tilde{\varepsilon}_Q)
    \Lambda(\tilde{\varepsilon}_Q) \, \mathrm{d}\tilde{\varepsilon}_Q \\
    & = \int_{0}^{\infty} \Lambda(\tilde{\varepsilon}_Q) \,
    G(\vec{p},\tilde{\varepsilon}_Q)\, \mathrm{d}\tilde{\varepsilon}_Q
    \,,
  \end{split}
  \label{eq42}
\end{equation}
where the ``distribution'' function $G(\vec{p},\tilde{\varepsilon_Q})$ is
defined as
\begin{equation}
  G(\vec{p},\tilde{\varepsilon}_Q) \equiv
  \int \delta(\varepsilon_Q-\tilde{\varepsilon}_Q) \mathrm{F}(\vec{P},T)
  \, \mathrm{d}^{3}P \,.
  \label{eq43}
\end{equation}
According to Ref.~\cite{fms89}, the function~(\ref{eq43}) is identical
with that defined by~Eq.~(\ref{eq6}). Using the definitions
(\ref{eq7})--(\ref{eq9}), the latter function can also be written down
in the following form:
\begin{equation}
  \begin{split}
    G(E,\varepsilon_Q) = & \sqrt{\frac{M_t}{\pi \mu_{t}T}} \,
    \frac{1}{\sqrt{E}} \exp
    \left[-\frac{M_{bx}}{T}\left(\frac{\varepsilon_Q}{\mu_t}
        +\frac{E}{M_{a\mu}}\right)\right] \\
    & \sinh \!
    \left(\frac{2M_{bx}}{T}\sqrt{\frac{\varepsilon_Q\, E}
        {\mu_{t} M_{a\mu}}} \, \right) .
  \end{split}
  \label{eq44}
\end{equation}
In particular, for $E\to{}0$, one has:
\begin{equation}
  G(E,\varepsilon_Q) = 2 \sqrt{\frac{1}{\pi T^3}
    \left( \frac{M_{bx}}{\mu_t} \right)^{\! 3}} \, \sqrt{\varepsilon_Q} \,
  \exp \! \left( -\frac{M_{bx}}{\mu_t}\frac{\varepsilon_Q}{T} \right) .
  \label{eq45}
\end{equation}
%

\section{Internal nuclear motion within  a hydrogenic  molecule}
\label{sec:nuc_motion}

Equation~(\ref{eq24}) can be expressed [see definitions~(\ref{eq9})
and~(\ref{eq11})] in terms of kinetic energies $\varepsilon_Q$ and
$\varepsilon_q$, which correspond to the momenta~$\vec{Q}$ and
$\vec{q}$, respectively:
\begin{equation}
  \begin{split}
    \varepsilon & = \mu\left(\frac{\varepsilon_Q}{\mu_t}\,
      + \beta_b^2\, \frac{\varepsilon_q}{\mu_{bx}} \,
      - 2 \beta_b \, \sqrt{\frac{\varepsilon_Q\,\varepsilon_q}
        {\mu_t\, \mu_{bx}}} \, z\right) , \\[5pt]
    z & = \cos\vartheta \,,
  \end{split}
  \label{eq46}
\end{equation}
where the angle between vectors $\vec{Q}$ and~$\vec{q}$ is denoted
by~$\vartheta$. When $\varepsilon_Q\gg\varepsilon_q$,
equation~(\ref{eq46}) takes a~simple asymptotic form~(\ref{eq10}):
$\varepsilon=(\mu/\mu_t)\varepsilon_Q$. On the other hand, when
$\varepsilon_Q\ll\varepsilon_q$, kinetic energy~$\varepsilon$ is mainly
determined by the internal molecular energy~$\varepsilon_q$. In
particular, at $\varepsilon_Q\to{}0$, the characteristic energy
$\varepsilon$ tends to a~constant value, which is given by the second
term of Eq.~(\ref{eq46}). The width of $\varepsilon$~spectrum is ruled
by the term $\sqrt{\varepsilon_Q\varepsilon_q}$ and thus rises with
increasing~$\varepsilon_Q$. On the other hand, as Eq.~(\ref{eq46})
indicates, a~fixed kinetic energy $\varepsilon_Q$~(\ref{eq9}) in CM of
the \textit{molecular} $a\mu+BX$ system corresponds to a~wide spectrum
of energies~$\varepsilon$ in the \textit{nuclear} CMS. As a~result, the
formation rate~$\Lambda(\varepsilon_Q)$ in the \textit{molecular} CMS,
which is given in~Eq.~(\ref{eq40}), has been obtained on averaging the
\textit{input} formation rates $\lambda(\varepsilon)$ over such
a~spectrum. This leads to an additional smearing of these rates (apart
from the thermal motion of the molecules), if they significantly change
within the spectrum of energies~$\varepsilon$. This problem has already
been considered in the case of $a\mu$ scattering from hydrogenic
molecules~\cite{adam06}, where it has been shown there that the
above-mentioned smearing effect is of importance for many low-energy
scattering cross sections $a\mu+BX$.

In order to obtain the distribution $g(\varepsilon_Q,\varepsilon)$ as
a~function of collision energies $\varepsilon$ in the \textit{nuclear}
CMS for a~fixed collision energy~$\varepsilon_Q$ in the
\textit{molecular}~CMS [see Eq.~(\ref{eq46})], we follow the calculating
scheme from Ref.~\cite{adam06}. It is assumed here that the orientation
of molecule with respect to the vector~$\vec{r}_m$ is random. In a
general case, the wave function~$\Phi_n(\vec{R}_0)$ of molecule~$BC$ in
the quantum state~$n$ can be written as follows:
\begin{equation}
  \label{eq:wave_mol_tot}
  \Phi_n(\vec{R}_0) = \frac{u_\nu(R_0)}{R_0}\,
  \mathrm{Y}_{KM_K}(\hat{\vec{R}}_0) \,,
  \qquad \hat{\vec{R}}_0 \equiv \frac{\vec{R}_0}{R_0} \,,
\end{equation}
where $n\equiv(\nu,K,M_K)$, the vibrational quantum number is denoted
by~$\nu$, the rotational state is specified by the quantum numbers $K$
and~$M_K$, and $\mathrm{Y}_{KM_K}$ stands for the corresponding
spherical harmonics. The radial wave function~$u_\nu$ in the harmonic
approximation takes the form
\begin{equation}
  \label{eq:wave_mol_rad}
  \begin{split}
    u_\nu(R_0) & = \mathcal{N}_\nu \mathrm{H}_\nu
    \bigl( \alpha\rho_0 \bigr) \,
    \exp \bigl( -\tfrac{1}{2}\alpha^2\rho_0^2 \bigr)\,, \\[5pt]
    \mathcal{N}_\nu & = \sqrt{\frac{\alpha}{2^\nu \nu! \, \sqrt{\pi}}}
    \,, \quad \alpha = \sqrt{\mu_{bx} \, \omega_0}
    \,, \quad \rho_0 =R_0 - \bar{R}_0 \,,
  \end{split}
\end{equation}
where $\mathrm{H}_\nu$ denotes the $\nu$-th Hermite polynomial and
$\rho_0$ is the displacement of $R_0$ from a~mean distance $\bar{R}_0$
between the nuclei $b$ and $x$ in~$BX$. The rotational~$E_K$ and
vibrational $E_\nu$ energy levels are given as
\begin{equation}
  \label{eq:energy_mol}
  E_K = B_\mathrm{rot} K(K+1) \,, \qquad
  E_\nu = (\nu+\tfrac{1}{2}) \, \omega_0 \,,
\end{equation}
where the rotational $B_\mathrm{rot}$ and vibrational~$\omega_0$
constants depend on the type of hydrogen isotopes $b$ and~$x$.

At temperatures usually applied in experiments, the molecules are in the
ground vibrational state $\nu=0$. Thus, the probability
density~(\ref{eq41}) for this state $(n=0)\equiv(\nu=0,K,M_K)$ takes the
form
\begin{equation}
  \label{eq:mom_mol_dis1}
  \begin{split}
  g(\varepsilon_Q,\tilde{\varepsilon})  = \int &
  \left| \int\exp(-i\vec{q}\cdot\vec{R}_0)\,
    \Phi_0(\vec{R}_0)\, \mathrm{d}^3R_0 \right|^2 \! \\
  & \delta(\varepsilon-\tilde{\varepsilon}) \,
  \frac{\mathrm{d}^3q}{(2\pi)^3} \,,
  \end{split}
\end{equation}
The substitution of the functions
(\ref{eq:wave_mol_tot})--(\ref{eq:wave_mol_rad}) into
Eq.~(\ref{eq:mom_mol_dis1}) results in
\begin{equation}
  \label{eq:mom_mol_dis2}
  g(\varepsilon_Q,\tilde{\varepsilon})  = \int
  f_{0K}(q)\, |\mathrm{Y}_{KM_K}(\hat{\vec{q}})|^2
  \delta(\varepsilon-\tilde{\varepsilon}) \, \mathrm{d}^3q \,,
\end{equation}
where
\begin{equation}
  \label{eq:f_kappa}
  f_{0K}(q)\equiv\frac{2\alpha}{\sqrt{\pi^3}}\, \mathcal{J}_R^2(q;K)
  \,, \qquad  \hat{\vec{q}} = \vec{q}/q \,,
\end{equation}
\begin{equation}
  \label{eq:I_R}
  \mathcal{J}_R(q;K)= \int_0^\infty \mathrm{j}_K(qR_0)
  \exp\left(-\tfrac{1}{2}\alpha^2 \rho_0^2 \right)
  R_0 \, \mathrm{d}R_0 \,,
\end{equation}
and $\mathrm{j}_K(qR_0)$ denotes the $K$-th spherical Bessel function.
The main contribution to the integral~(\ref{eq:I_R}) comes from the
vicinity of $\,\rho_0\approx 0\, (R_0\approx\bar{R}_0)$. Thus, we obtain
a~good approximation of~$\mathcal{J}_R$ if the lower integration limit
is extended to~$-\infty$. Then, we can use the asymptotic form of the
function~$\mathrm{j}_K(x)$ for $x\gg{}K(K+1)$~\cite{abramowitz}
\begin{equation}
  \label{eq:bessel_asym}
  \mathrm{j}_K(x)\approx
  \frac{1}{x}\cos\!\left[x-\tfrac{1}{2}(K+1)\pi\right] \,,
\end{equation}
which is the exact form for $K=0$ at any~$x$. As a~result, we obtain the
following approximation:
\begin{equation}
  \label{eq:I_R_approx}
   \mathcal{J}_R(q;K) \approx \sqrt{2\pi}\, \frac{\bar{R}_0}{\alpha}\,
   \mathrm{j}_K(q\bar{R}_0) \exp\left(-\frac{q^2}{2\alpha^2}\right).
\end{equation}
In the case of $K>0$ and $q\bar{R}_0\lesssim{}K(K+1)$, the integral
$\mathcal{J}_R$ should be estimated numerically.

Further evaluation of Eq.~(\ref{eq:mom_mol_dis2}) for
$g(\varepsilon_Q,\tilde{\varepsilon})\equiv %
g_K(\varepsilon_Q,\tilde{\varepsilon})$
is performed using the additional averaging over the projections $M_K$
with a~uniform weight $(2K+1)^{-1}$ for a~fixed~$K$
\begin{equation}
  \label{eq:gdis1}
  \begin{split}
    g_K(&\varepsilon_Q,\tilde{\varepsilon})  =\frac{1}{2K+1}\sum_{M_K}
    \int f_{0K}(q)\, |\mathrm{Y}_{KM_K}(\hat{\vec{q}})|^2
    \delta(\varepsilon-\tilde{\varepsilon})\, \mathrm{d}^3q \\[5pt]
    & = \int_0^\infty f_{0K}(q)\, q^2 \, \mathrm{d}q
    \int_0^{2\pi} \frac{1}{2K+1} \sum_{M_K}
    \mathcal{J}_{KM_K}(\varepsilon_q) \, \mathrm{d}\phi \,.
  \end{split}
\end{equation}
Here, the integral $\mathcal{J}_{KM_K}(\varepsilon_q)$ is specified as
\begin{equation}
  \label{eq:I_KM}
  \mathcal{J}_{KM_K}(\varepsilon_q) = \int_{0}^{\pi}
  |\mathrm{Y}_{KM_K}(\vartheta,\phi)|^2
  \delta(\varepsilon-\tilde{\varepsilon}) \sin\vartheta\,
  \mathrm{d}\vartheta
\end{equation}
and the solid angle $\Omega(\vartheta,\phi)$ determines the orientation
of vector $\vec{q}$ with respect to a~fixed vector~$\vec{Q}$:
$\mathrm{d}\Omega=\sin\vartheta\,\mathrm{d}\vartheta\,\mathrm{d}\phi%
=\mathrm{d}z\,\mathrm{d}\phi$.
Then, on using the relation~(\ref{eq46}), Eq.~(\ref{eq:I_KM}) can be
expressed as the following integral
\begin{equation}
  \label{eq:int_J1}
  \begin{split}
  \mathcal{J}_{KM_K} =
  \int_{-1}^1 & |\mathrm{Y}_{KM_K}(\vartheta,\phi)|^2 \,
  \delta \left(\frac{\mu}{\mu_t}\, \varepsilon_Q
   + \beta_b^2\, \frac{\mu}{\mu_{bx}}\, \varepsilon_q  \right. \\[5pt]
   & \left. -\tilde{\varepsilon}
    - 2 \beta_b \, \frac{\mu}{\sqrt{\mu_t\, \mu_{bx}}} \,
    \sqrt{\varepsilon_Q \, \varepsilon_q}\, z \right) \mathrm{d}z \,
  \end{split}
\end{equation}
with respect to the variable~$z$. The delta function impose the
following condition on $z$:

\begin{equation}
  \label{eq:z0_cond}
  z=z_0\equiv\cos\vartheta_0 \,, \quad
  z_0 = \frac{1}{2\beta_b}\, \frac{\sqrt{\mu_t\, \mu_{bx}}}{\mu}
  \frac{1}{\sqrt{\varepsilon_Q\, \varepsilon_q}}
  \left( \frac{\mu}{\mu_t}\, \varepsilon_Q
   +\beta_b^2\frac{\mu}{\mu_{bx}}\,\varepsilon_q -\tilde{\varepsilon},
   \right),
\end{equation}
for all $z_0$ such that $|z_0|\leq 1$. The result of the integration
over~$z$ can be written down as
\begin{equation}
  \label{eq:int_J2}
  \begin{split}
    \mathcal{J}_{KM_K} = & \frac{1}{2\beta_b}
    \frac{\sqrt{\mu_t\, \mu_{bx}}}{\mu} \,
    \frac{1}{\sqrt{\varepsilon_Q\, \varepsilon_q}} \,
    \Theta(\varepsilon_q-\varepsilon_{Q_1}) \\
    & \Theta(\varepsilon_{Q_2}-\varepsilon_q) \,
    |\mathrm{Y}_{KM_K}(\vartheta_0,\phi)|^2  \,,
  \end{split}
\end{equation}
where $\Theta$ is the Heaviside step function and
\begin{equation}
  \label{eq:roots_cond}
  \varepsilon_{Q_{1,2}} = \frac{1}{\beta_b^2}
  \frac{\mu_{bx}}{\mu\,\mu_t}
  \left(\sqrt{\mu_t\, \tilde{\varepsilon}}\, \mp
    \sqrt{\mu\, \varepsilon_Q} \right)^2 .
\end{equation}
Substituting Eq.~(\ref{eq:int_J2}) into Eq.~(\ref{eq:gdis1}) and
changing the integration variable $q$ to variable $\varepsilon_q$, which
is defined in Eq.~(\ref{eq11}), lead to the following relation
\begin{equation}
  \label{eq:gdis3}
  \begin{split}
    g_K(\varepsilon_Q,\tilde{\varepsilon}) = & \frac{1}{2\beta_b}
    \frac{\sqrt{\mu_t\, \mu_{bx}}}{\mu} \, \frac{1}{\sqrt{\varepsilon_Q}}
    \int_{\varepsilon_{Q_1}}^{\varepsilon_{Q_2}}
    f_{0K}(\varepsilon_q) \, \mathrm{d}\varepsilon_q \\[5pt]
    & \int_0^{2\pi} \frac{1}{2K+1} \sum_{M_K}
    |\mathrm{Y}_{KM_K}(\vartheta_0,\phi)|^2 \, \mathrm{d}\phi \,,
  \end{split}
\end{equation}
where
\begin{equation}
  \label{eq:feq}
  f_{0K}(\varepsilon_q) \equiv \sqrt{2\mu_{bx}^3}\,
  f_{0K}(q(\varepsilon_q))\,.
\end{equation}
The integration over $\phi$ in Eq.~(\ref{eq:gdis3}) gives the factor
of~1/2, since for any~$\Omega$ one has
\begin{equation}
  \label{eq:sum_harmonics}
   \frac{1}{2K+1} \sum_{M_K}\bigl| \mathrm{Y}_{KM_K}(\Omega)\bigr|^2
   = \frac{1}{4\pi} \,.
\end{equation}
Thus, for a~random-oriented molecule,
\begin{equation}
  \label{eq:gdis4}
  g_K(\varepsilon_Q,\tilde{\varepsilon}) = \frac{1}{4\beta_b}
  \frac{\sqrt{\mu_t\, \mu_{bx}}}{\mu} \, \frac{1}{\sqrt{\varepsilon_Q}}
  \int_{\varepsilon_{Q_1}}^{\varepsilon_{Q_2}} f_{0K}(\varepsilon_q) \,
  \mathrm{d}\varepsilon_q \,.
\end{equation}
Employing a convenient variable
\begin{equation}
  \label{eq:omega}
  \omega = 2\varepsilon_q/\omega_0
\end{equation}
and Eqs.~(\ref{eq11}), (\ref{eq:wave_mol_rad}) and (\ref{eq:feq}), the
distribution~(\ref{eq:gdis4}) takes the final form
\begin{equation}
  \label{eq:gdis6}
  \begin{split}
    g_K(\varepsilon_Q,\tilde{\varepsilon}) & =
    \frac{\alpha^4}{\beta_b\sqrt{(2\pi)^3}}
    \frac{\sqrt{\mu_t\, \mu_{bx}}}{\mu}\,
    \frac{1}{\sqrt{\omega_0\varepsilon_Q}} \\[5pt]
    & \int_{\omega_1}^{\omega_2} \mathcal{J}_R^2
    (\alpha\sqrt{\omega};K)\, \mathrm{d}\omega \,,
    \quad \omega_{1,2} \equiv \frac{2\varepsilon_{Q_{1,2}}}{\omega_0}
    \,,
  \end{split}
\end{equation}
for a~fixed collision energy~$\varepsilon_Q$ in the
\textit{molecular}~CMS.

Let us consider the limit $\varepsilon_Q\to{}0$. Then, according to
Eq.~(\ref{eq46}), the corresponding distribution
of~$\tilde{\varepsilon}$ is solely determined by the distribution of
internal kinetic energy~$\varepsilon_q$ of the molecule~$BX$:
\begin{equation}
  \label{eq:limit_cond_small}
  \tilde{\varepsilon}\sim\varepsilon_q\gg\varepsilon_Q  \,.
\end{equation}
In this approximation
\begin{equation}
  \label{eq:limits_approx}
  \begin{split}
    & \omega_{1,2} \approx \tilde{\omega} = \frac{2}{\beta_b^2}
    \frac{\mu_{bx}}{\mu}\frac{\tilde{\varepsilon}}{\omega_0} \,, \\
    & \mathrm{d}\omega \to \Delta\omega = \omega_2 - \omega_1 =
    \frac{8}{\beta_b^2} \frac{\mu_{bx}}{\sqrt{\mu_t\mu}}
    \frac{\sqrt{\varepsilon_Q\tilde{\varepsilon}}}{\omega_0} \,,
  \end{split}
\end{equation}
so that
\begin{equation}
  \label{eq:f_eb_eps0}
  g_K(\varepsilon_Q,\tilde{\varepsilon}) =
  \frac{2\alpha^4}{\beta_b^2 \sqrt{\pi^3}}\,
  \frac{\mu_{bx}}{\mu}\, \frac{1}{\omega_0} \sqrt{\tilde{\omega}}\;
  \mathcal{J}_R^2(\alpha\sqrt{\tilde{\omega}};K) \,,
\end{equation}
when $\varepsilon_Q\to 0$. In this limit, if the
approximation~(\ref{eq:I_R_approx}) is valid, Eqs.~(\ref{eq:gdis6})
and~(\ref{eq:f_eb_eps0}) take the following asymptotic forms:
\begin{equation}
  \label{eq:gdis_approx}
  \begin{split}
    g_K(\varepsilon_Q,\tilde{\varepsilon}) = &
    \frac{\alpha^2 \bar{R}_0^2}{\beta_b\sqrt{2\pi}}
    \frac{\sqrt{\mu_t\, \mu_{bx}}}{\mu}\,
    \frac{1}{\sqrt{\omega_0\varepsilon_Q}} \\[5pt]
    & \int_{\omega_1}^{\omega_2} \mathrm{j}_K^2
    (\alpha\bar{R}_0\sqrt{\omega})\exp(-\omega)\, \mathrm{d}\omega
  \end{split}
\end{equation}
and
\begin{equation}
  \label{eq:f_eb_eps0_approx}
  g_K(\varepsilon_Q,\tilde{\varepsilon}) =
  \frac{4\alpha^2\bar{R}_0^2}{\beta_b^2 \sqrt{\pi}}\,
  \frac{\mu_{bx}}{\mu}\, \frac{1}{\omega_0} \sqrt{\tilde{\omega}}\;
  \mathrm{j}_K^2(\alpha\bar{R}_0\sqrt{\tilde{\omega}})
  \exp(-\tilde{\omega}) \,,
\end{equation}
respectively.

In the limit $\varepsilon_Q\to\infty$, the argument of the
$\delta$~function in~Eq.~(\ref{eq:int_J1}) tends to
$\tilde{\varepsilon}-\mu\,\varepsilon_Q/\mu_t$, which means that
\begin{equation}
  \label{eq:limit_eps_infty}
  \tilde{\varepsilon} \to \frac{\mu}{\mu_t}\, \varepsilon_Q \,,\quad
  \mathrm{when} \quad \varepsilon_Q\to\infty \,.
\end{equation}
Let us note once more that it is invalid to use the above asymptotic
relation, which is equivalent to Eq.~(\ref{eq10}), when the condition
$\varepsilon_Q\gg \varepsilon_q\sim\omega_0/4$ is not fulfilled.

\section{Results and discussion}
\label{sec:discussion}

The distribution density~(\ref{eq:gdis6}), which was calculated using
a~numerical integration in the case of $p\mu+\mathrm{H}_2$ system,
\begin{figure}[htb]
  \centering
  \includegraphics[width=8cm]{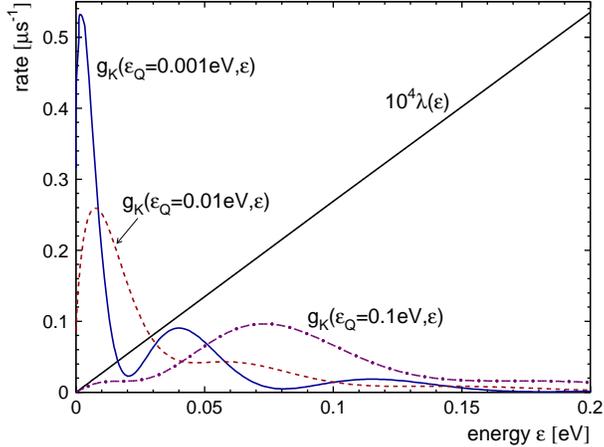}
  \caption{(Color online) Function $g_{K=0}(\varepsilon_Q,\varepsilon)$
    for $p\mu+\mathrm{H}_2$ at $\varepsilon_Q=0.001$, 0.01, and 0.1~eV
    (in arbitrary units), together with the rate $\lambda(\varepsilon)$
    of nonresonant $pp\mu$ formation in the transition
    ${J=1\to{}J=1,~v=0}$.}
  \label{fig:enucdis}
\end{figure}%
is shown\footnote{For a simplicity of the notation, the integration
  variable $\tilde{\varepsilon}$ in the presented plots of the
  functions~(\ref{eq:gdis6}) and~(\ref{eq:f_eb_eps0}) is displayed
  without the ``tilde'' sign.}%
in Fig.~\ref{fig:enucdis} as a~function of kinetic energy $\varepsilon$
in the \textit{nuclear} CMS for several values of~$\varepsilon_Q$ (in
the \textit{molecular} CMS).  Since the characteristic kinetic
energy~$\omega_{bx}$ in the molecular ground state is determined by the
zero-point vibrations:
$\omega_{bx}\sim\tfrac{1}{2}E_{\nu=0}=\omega_0/4$, the magnitude of
$\omega_{bx}$ for the H$_2$ molecule is on the order of 0.1~eV.
Therefore, the curve plotted in Fig.~\ref{fig:enucdis} for
$\varepsilon_Q=0.001$~eV practically represents the
limit~(\ref{eq:f_eb_eps0}). It is evident that the averaging over the
distribution density $g\equiv{}g_K$ in Eq.~(\ref{eq40}) is important
when the rate $\lambda(\varepsilon)$ significantly changes within the
characteristic width of~$g_K$, which occurs in the case of $pp\mu$
formation presented in this figure.

\begin{figure}[htb]
  \centering
  \includegraphics[width=8cm]{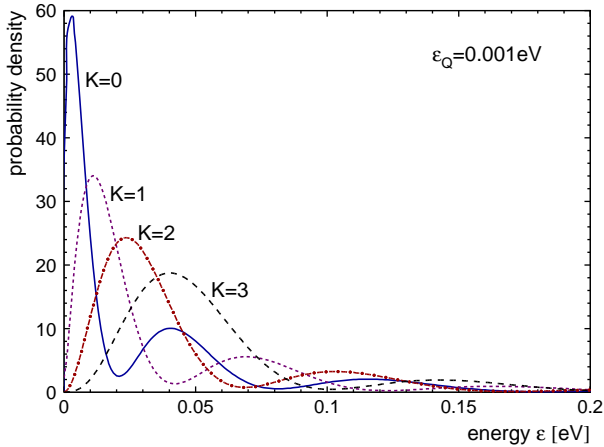}
  \caption{(Color online) The distribution density
    $g_K(\varepsilon_Q=0.001\,\mathrm{eV},\varepsilon;K)$ for various
    rotational quantum numbers~$K$ in the case of $p\mu+\mathrm{H}_2$
    process.}
  \label{fig:enucdis_001_K}
\end{figure}

\begin{figure}[htb]
  \centering
  \includegraphics[width=8cm]{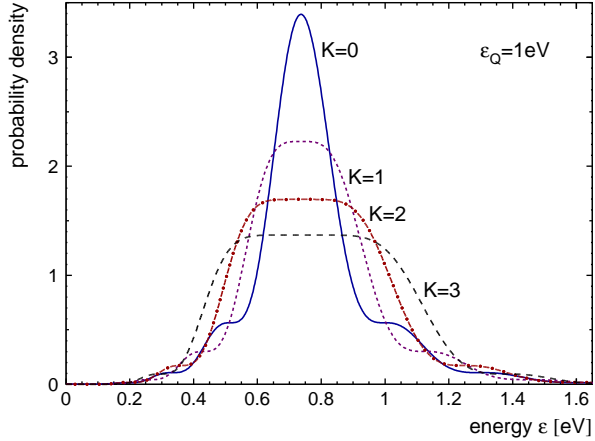}
  \caption{(Color online) The same as in Fig.~\ref{fig:enucdis_001_K}
    for $\varepsilon_Q=1$~eV.}
  \label{fig:enucdis_1_K}
\end{figure}

The dependence of~$g_K$ on the rotational quantum numbers~$K$ is shown
in Figs.~\ref{fig:enucdis_001_K} and~\ref{fig:enucdis_1_K} for
$\varepsilon_Q=0.001$ and~1~eV, respectively. At
$\varepsilon_Q\lesssim{}0.001$~eV, the distribution of~$\varepsilon$ is
practically determined by the rotational-vibrational state of the H$_2$
molecule.
However, this distribution strongly changes for subsequent rotational
numbers. The distribution maximum decreases with rising~$K$. At
$\varepsilon_Q\gtrsim{}1$~eV, the location of maximum of~$g_K$ is
proportional to~$\varepsilon_Q$, according to the asymptotic
relation~(\ref{eq:limit_eps_infty}). The rotational-vibrational motion
of the molecule causes a~significant broadening of this maximum, which
is wider at higher~$\varepsilon_Q$. The distribution~$g_K$ is flatter
for greater~$K$. The presented changes of~$g_K$ for various~$K$ can
lead to an appreciable dependence of the formation rates~(\ref{eq39})
and~(\ref{eq40}) on the initial rotational distribution of the target
molecules.

In order to demonstrate effects of the internal nuclear motion in the
target molecules, the \textit{molecular} formation rate
$\Lambda(\varepsilon_Q(\varepsilon))$, which was calculated using
Eq.~(\ref{eq40}), and the input \textit{nuclear} formation rate
$\lambda(\varepsilon)$ are plotted below for some interesting cases. For
the sake of comparison, the pairs of the corresponding rates
$\Lambda(\varepsilon_Q(\varepsilon))$ and $\lambda(\varepsilon)$ are
shown together using the asymptotic relation~(\ref{eq:limit_eps_infty})
between kinetic energies $\varepsilon_Q$ and~$\varepsilon$. Since we
consider target temperatures $T\leq{}300$~K, the hydrogenic molecules
are always in the ground vibrational state~$\nu=0$

\begin{figure}[htb]
  \centering
  \includegraphics[width=8cm]{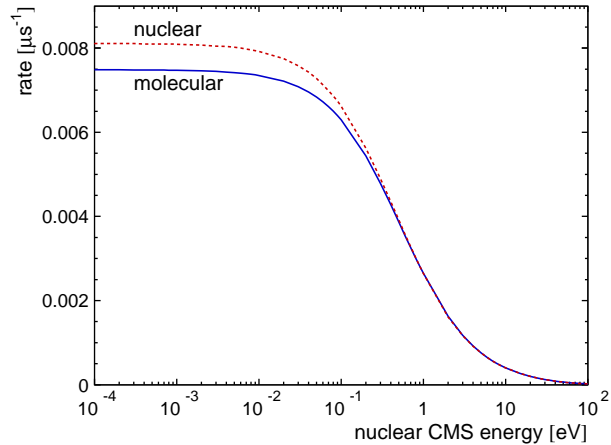}
  \caption{(Color online) The rates $\Lambda$ (\textit{molecular}) and
    $\lambda$ (\textit{nuclear}) of nonresonant $pp\mu$ formation versus
    kinetic energy~$\varepsilon$, for the transition $J=0\to{}J=0$ in
    collision $p\mu+\mathrm{H}_2(K=0)$.}
  \label{fig:nucmov_ppm_J0_J0}
\end{figure}

\begin{figure}[htb]
  \centering
  \includegraphics[width=8cm]{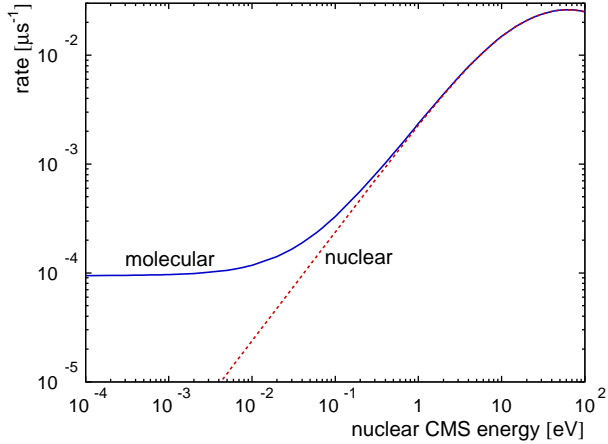}
  \caption{(Color online) The same as in Fig.~\ref{fig:nucmov_ppm_J0_J0}
    for the transition $J=1\to{}J=0$.}
  \label{fig:nucmov_ppm_J1_J0}
\end{figure}

The internal-motion effect in $pp\mu$ formation for the transition
$J=0\to{}J=0$, $v=0$ is shown in~Fig.~\ref{fig:nucmov_ppm_J0_J0}.
The apparent difference between the rates below about 0.2~eV is due to
a~significant variation of the input rate $\lambda(\varepsilon)$ within
a~relatively narrow interval
5~meV~$\lesssim\varepsilon\lesssim{}0.2$~eV, which is comparable with
the characteristic width of the distribution~$g_K$. This effect
disappears at higher energies. The nuclear-motion smearing of the
formation rate for the $pp\mu$ formation in the transition $J=1\to{}J=0$
is shown in Fig.~\ref{fig:nucmov_ppm_J1_J0}.
Here, this effect is due to the small magnitude of
$\lambda(\varepsilon)$ for the $p\mu+p$ scattering at
$\varepsilon\to{}0$ in the initial $J=1$ state. In general, such
a~behavior is typical for the nonresonant formation at the muonic atom
scattering in higher partial ($J>0$) wave state.

\begin{figure}[htb]
  \centering
  \includegraphics[width=8cm]{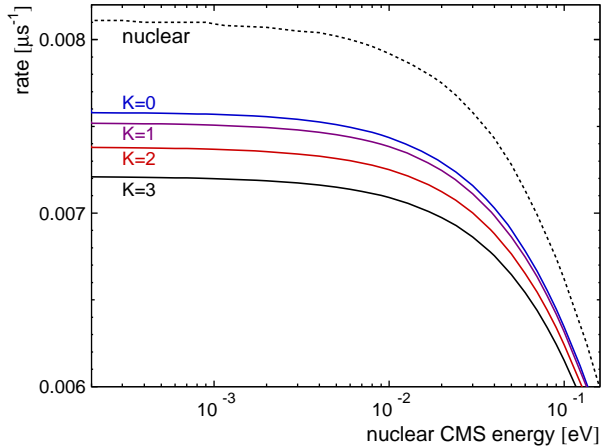}
  \caption{(Color online) The same as in
    Fig.~\ref{fig:nucmov_ppm_J0_J0}. The \textit{molecular}
    rates~$\Lambda$ (solid lines) are calculated for the initial
    rotational states $K=0$, 1, 2, and~3 of the H$_2$ molecule.}
  \label{fig:rotrate_ppm_J0_J0}
\end{figure}

\begin{figure}[htb]
  \centering
  \includegraphics[width=8cm]{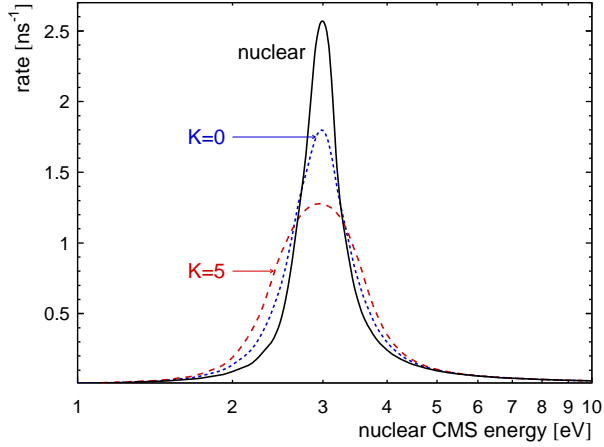}
  \caption{(Color online) The \textit{molecular} $\Lambda$ (for $K=0$
    and 5, dashed lines) rates and the \textit{nuclear} $\lambda$ rate
    for the nonresonant $tt\mu$ formation in the transition
    $J=2\to{}J=2$.}
  \label{fig:rotrate_ttm_J2_J2}
\end{figure}

A~dependence of the formation rate~$\Lambda$ on the initial rotational
state of the target molecule is shown in
Fig.~\ref{fig:rotrate_ppm_J0_J0},
for the $pp\mu$ formation in the transition $J=0\to{}J=0$. The
appreciable differences of~$\Lambda$ for various~$K$ lead to different
values of the average formation rate $\lambda(E,T)$ in the LAB frame,
depending on the rotational population for a~specified target and
temperature. For example, strong nuclear-motion and rotational effects
occur at relatively high energies (target temperatures) for the
transition $J=2\to{}J=2$ in the nonresonant $tt\mu$ formation within the
T$_2$ molecule (see Fig.~\ref{fig:rotrate_ttm_J2_J2}).
This is due to the existence of the $tt\mu$ quasi-stationary state,
which appears as a narrow peak in the elastic $t\mu+t$ cross-section
at the scattering energy $\sim 3$~eV and $J=2$~\cite{ppp75,bf87}.

\begin{figure}[htb]
  \centering
  \includegraphics[width=8cm]{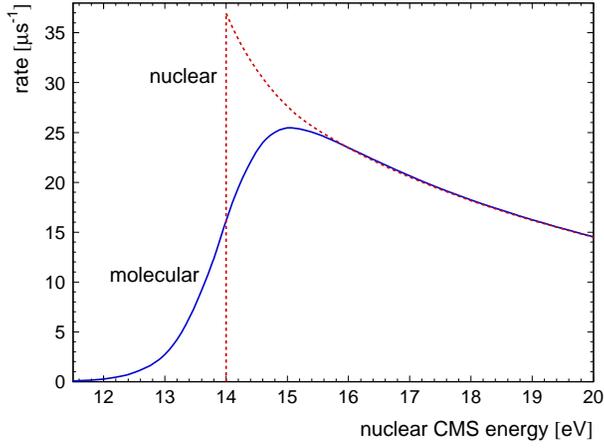}
  \caption{(Color online) The total \textit{molecular} $\Lambda$ (solid
    line) and \textit{nuclear} $\lambda$ (dashed line) rates of
    nonresonant $dd\mu$ formation in the $J=\upsilon=1$ state versus
    energy~$\varepsilon$. The Boltzmann population of the rotational
    energy levels of~D$_2$ for $T=300$~K is assumed.}
  \label{fig:nucmov_ddm_Jv11}
\end{figure}

\begin{figure}[htb]
  \centering
  \includegraphics[width=8cm]{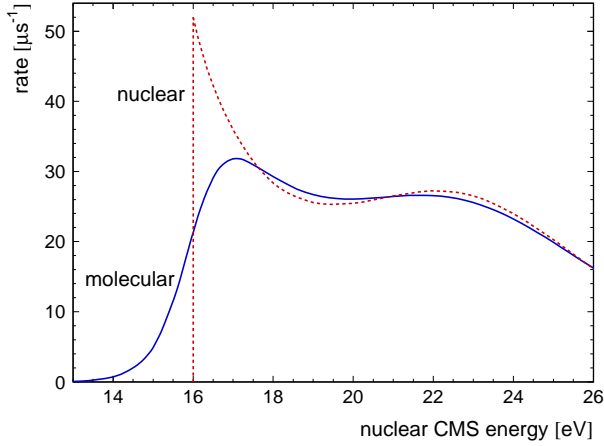}
  \caption{(Color online) The same as in Fig.~\ref{fig:nucmov_ddm_Jv11}
    for the $dt\mu$ formation in the state $J=\upsilon=1$ in collision
    $t\mu+\mathrm{D}_2$.}
  \label{fig:nucmov_tdm_Jv11}
\end{figure}

Figure~\ref{fig:nucmov_ddm_Jv11} refers to the nonresonant formation of
the $dd\mu$ molecule~(\ref{eq2}) in the loosely-bound state
$J=\upsilon=1$ (see details in Ref.~\cite{af12}).
For the thermalized $d\mu$ atoms, the nonresonant process~(\ref{eq2}) is
unfeasible. However, such a~reaction is effective in the case of
nonthermalized $d\mu$ atoms with collision energies exceeding the
ionization threshold of the D$_2$ molecule. The rate
$\Lambda(\varepsilon_Q(\varepsilon))$ plotted in this figure was
estimated by averaging the corresponding input rate
$\lambda(\varepsilon)$ for $dd\mu$ formation in the $J=\upsilon=1$
state~\cite{af12}, with the use of Eqs.~(\ref{eq40})
and~(\ref{eq:gdis6}). A~smearing of the rate~$\Lambda$ due to the
internal nuclear motion is quite strong near the threshold. However,
this effect cannot significantly affect the kinetics of $dd$-fusion,
since the threshold is located at energy much higher than the thermal
energies of typical D$_2$ targets. Thus, a~fraction of the
nonthermalized $d\mu$ atoms with such high energies ($\sim{}15$~eV) is
small. The analogous around-threshold smearing effect is shown in
Fig.~\ref{fig:nucmov_tdm_Jv11} for nonresonant formation of the $dt\mu$
molecule in the loosely-bound state $J=\upsilon=1$.
This muonic molecule is created in the collision of $t\mu$ with the
D$_2$ molecule. In this case, the threshold energy of 16~eV is somewhat
greater than that for the $dd\mu$ molecule.

\begin{figure}[htb]
  \centering
  \includegraphics[width=8cm]{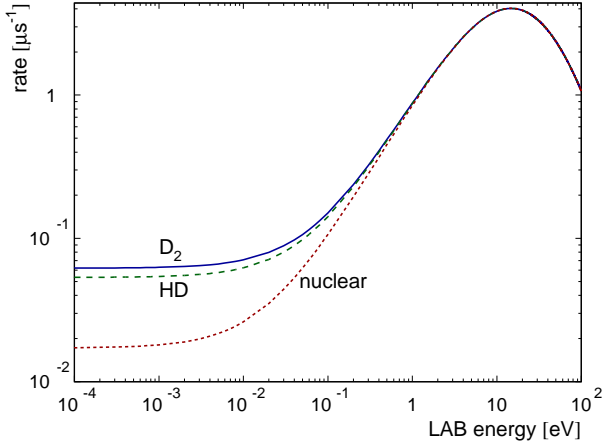}
  \caption{(Color online) The rate $\Lambda(E)$ of nonresonant $dd\mu$
    formation in the $J=0$ state in $d\mu$ collision with the target
    molecules D$_2$ and HD versus LAB energy~$E$. The molecules have the
    Boltzmann populations of their rotational levels for 30~K.}
  \label{fig:nucmov_ddm_dp_30K}
\end{figure}

The \textit{molecular} rates $\Lambda(E)$ of nonresonant $dd\mu$
formation in the state $J=0$, in the case of $d\mu$ collision with the
molecules D$_2$ and~HD, are presented in
Fig.~\ref{fig:nucmov_ddm_dp_30K} as functions of LAB kinetic energy~$E$.
These rates were calculated using Eqs.~(\ref{eq42}) and~(\ref{eq44}).
The corresponding \textit{nuclear} rate $\lambda(E)$ is plotted for
a~comparison. A~strong smearing of $\Lambda(E)$ appears at low collision
energies ($E\lesssim{}0.1$~eV), where the input
rate~$\lambda(\varepsilon)$ rises rapidly in the interval of energies
$\varepsilon\ll\omega_0$. As a~result, the molecular rates are much
greater than the corresponding nuclear rate for thermalized $d\mu$
atoms. Also, an appreciable isotopic effect can be observed in
Fig.~\ref{fig:nucmov_ddm_dp_30K} at low energies. This effect is due to
the different distributions $g(\varepsilon_Q,\varepsilon)$ of the
deuteron kinetic energy in the molecules D$_2$ and~HD. Since most
kinetic energy in the HD molecule is carried by the lighter proton, it
follows that the mean kinetic energy of deuteron in~HD is smaller than
that in the D$_2$ molecule. Let us note that the nonresonant $dd\mu$
formation rate in the state ($J=1$, $\upsilon=0$) is very flat at low
collision energies.  As a~result, there is no appreciable differences
between the rates $\lambda(E)$ and $\Lambda(E)$ in this case.

\begin{figure}[htb]
  \centering
  \includegraphics[width=8cm]{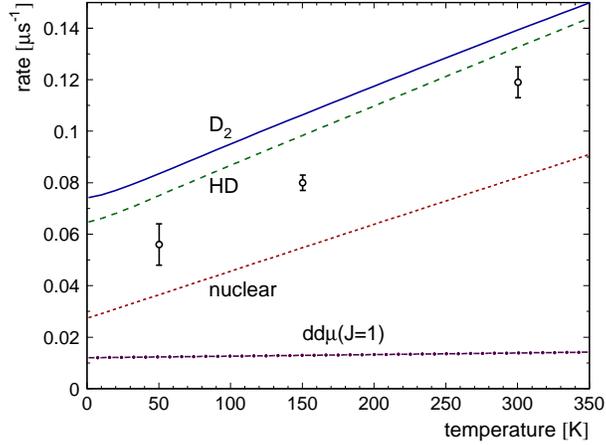}
  \caption{(Color online) The thermally averaged total rates
    $\lambda(T)$ and $\Lambda(T)$ (for D$_2$ and HD) of nonresonant
    $dd\mu$ formation as functions of the target temperature~$T$. The
    experimental data for the HD target are taken from
    Ref.~\cite{bal11}. A~contribution from the $J=1$ state is also
    plotted (dash-dotted line).}
  \label{fig:nres_ddm_temp}
\end{figure}

Significant differences between the rates $\lambda(E)$ and~$\Lambda(E)$
at lowest energies are of special importance when the conditions of
steady-state kinetics are reached, e.g., as in the case of experiments
reported in Ref.~\cite{bal11}. Under such conditions, the kinetics is
described in terms of the thermally averaged rates of various $\mu$CF
processes for a~fixed target temperature. In calculations it is assumed
that the distributions of LAB kinetic energies of the molecules and
muonic atoms in a~gaseous target have the Maxwellian shape. The
rotational levels of the target molecules obey the Boltzmann
distribution. The thermally-averaged total nonresonant rates
$\lambda(T)$ and $\Lambda(T)$ are plotted in
Fig.~\ref{fig:nres_ddm_temp}
as functions of the target temperature, for the D$_2$ and HD gases.
Also, the contribution from the state $J=1$ to the total rate is shown,
which is practically identical for both the nuclear and molecular rates.
The experimental data points at $T=50$, 150, and 300~K were determined
in Ref.~\cite{bal11} using the steady-state kinetics for a~pure HD
target. The calculated rates $\lambda(T)$ and $\Lambda(T)$ do not well
describe the data. Nevertheless, the molecular rate $\Lambda(T)$ for the
HD target, which takes into account the deuteron motion within the
molecule HD, is closer to the data than the nuclear rate $\lambda(T)$.
From Fig.~\ref{fig:nres_ddm_temp} one can conclude that the theoretical
rate~$\Lambda(T)$, which takes into account an appreciable isotopic
effect, is closer to the experimental data (compare the rates for D$_2$
and~HD.

\begin{figure}[htb]
  \centering
  \includegraphics[width=8cm]{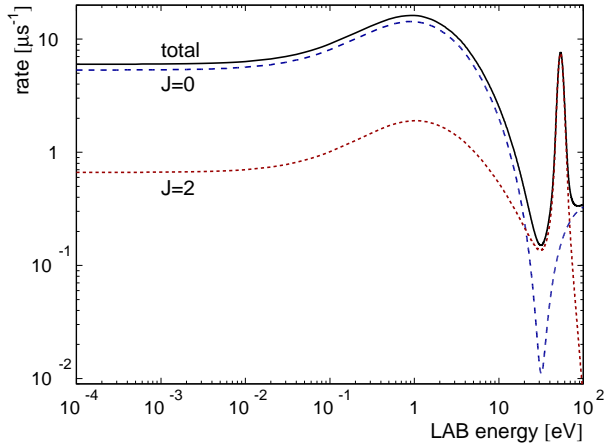}
  \caption{(Color online) The contributions from the rotational states
    $J=0$ and~2 to the total molecular rate $\Lambda(E)$ of nonresonant
    $dt\mu$ formation in~D$_2$ at 300~K.}
  \label{fig:nucmov_tdm_300K}
\end{figure}%
The contributions to the total rate $\Lambda(E)$ from the nonresonant
$dt\mu$ formation in the rotational states~$J=0$ and~2 are shown in
Fig.~\ref{fig:nucmov_tdm_300K} as functions of LAB energy.  For the sake
of clarity, a~higher-energy contribution from the state $J=\upsilon=1$
of $dt\mu$, which is plotted in Fig.~\ref{fig:nucmov_tdm_Jv11}, is not
included here. A~negligible contribution from the state $J=1$,
$\upsilon=0$ of $dt\mu$ is not visible in this plot. The presented rates
have been calculated assuming the 300-K Boltzmann distribution of the
rotational levels of the target D$_2$ molecules.

Our calculations of the thermally-averaged molecular rates $\Lambda(T)$
are summarized in Tables \ref{tab:rate_K} and~\ref{tab:rate_J}. 
\begin{table*}[htb]
  \begin{center}
    \caption {The calculated averaged nonresonant formation rates of
      muonic molecules~[$10^6$~s$^{-1}$] for different populations of
      the rotational numbers~$K$ of the target hydrogenic molecules.}
    \label{tab:rate_K}
    \begin{ruledtabular}
      \newcolumntype{.}{D{.}{.}{1.4}}
      \begin{tabular}{cc....}
        \multicolumn{1}{c}{Muonic} & \multicolumn{1}{c}{Target}
        &\multicolumn{3}{c} {Temperature $T=30$~K}
        & \multicolumn{1}{c}{Temperature $T=300$~K} \\
        \multicolumn{1}{c}{molecule} & \multicolumn{1}{c}{molecule}
        & \multicolumn{1}{c}{$K=0$} & \multicolumn{1}{c}{$K=1$}
                                     & \multicolumn{1}{c}{Statistical}
        & \multicolumn{1}{c}{Boltzmann distribution} \\
        \hline
        $pp\mu$ & H$_2$ & 1.806 & 1.805 & 1.805 & 1.799 \\
        $pd\mu$ & H$_2$ & 5.626 & 5.618 & 5.620 & 5.582 \\
        $pt\mu$ & H$_2$ & 6.375 & 6.363 & 6.366 & 6.317 \\
        $dd\mu$ & D$_2$ & 0.0788 & 0.0829 & 0.0801 & 0.139 \\
        $dt\mu$ & D$_2$ & 4.083 & 4.433 & 4.200 & 7.249 \\
        $tt\mu$ & T$_2$ & 2.685 & 2.681 & 2.682 & 2.639 \\
      \end{tabular}
    \end{ruledtabular}
  \end{center}
\end{table*}%
The rates for the rotational states $K=0$,~1 of target molecules at 30~K
are separately shown in Table~\ref{tab:rate_K}. The corresponding rate
for the statistical mixture of these states is also given.
\begin{table*}[htb]
  \begin{center}
    \caption {The calculated averaged nonresonant formation rates
      [$10^6$~s$^{-1}$] for different rotational numbers~$J$ of the
      muonic-molecule states~$Jv$. The Boltzmann distribution of
      rotational states of the target molecules is assumed for the both
      temperatures.}
    \label{tab:rate_J}
    \newcolumntype{.}{D{.}{.}{1.4}}
    \begin{ruledtabular}
      \begin{tabular}{cc......}
        \multicolumn{1}{c}{Temperature} & \multicolumn{1}{c}{State} &
        \multicolumn{6}{c}{Muonic molecule (Target molecule)} \\
        \multicolumn{1}{c}{[K]} & \multicolumn{1}{c}{$J$} &
        \multicolumn{1}{c}{$pp\mu$ (H$_2$)} &
        \multicolumn{1}{c}{$dd\mu$ (HD)}    &
        \multicolumn{1}{c}{$dd\mu$ (D$_2$)} &
        \multicolumn{1}{c}{$dt\mu$ (D$_2$)} &
        \multicolumn{1}{c}{$dt\mu$ (DT)}    &
        \multicolumn{1}{c}{$tt\mu$ (T$_2$)} \\
        \hline\noalign{\vskip2pt}
    & 0     & 0.008 & 0.0568 & 0.0653 & 3.655 & 3.893 & 0.003 \\
30  & 1     & 1.798 & 0.0119 & 0.0119 & 0.000 & 0.000 & 2.680 \\
    & 2 & \multicolumn{1}{c}{---} & 0.0015 & 0.0019 & 0.454 & 0.484 & 0.000\\
    & total & 1.806 & 0.0702 & 0.0791 & 4.109 & 4.377 & 2.683 \\
\hline
    & 0     & 0.007 & 0.1170 & 0.1234 & 6.444 & 6.550 & 0.003 \\
300 & 1     & 1.792 & 0.0118 & 0.0118 & 0.000 & 0.000 & 2.636 \\
    & 2 & \multicolumn{1}{c}{---} & 0.0038 & 0.0041 & 0.805 & 0.819 & 0.000\\
    & total & 1.799 & 0.1326 & 0.1393 & 7.249 & 7.369 & 2.639 \\
      \end{tabular}
    \end{ruledtabular}
  \end{center}
\end{table*}%
These particular rates were calculated since at low temperatures the
rotational levels of hydrogen-isotope molecules in certain experimental
targets are not equilibrated according to the Boltzmann distribution. On
the other hand, at $T\gg{}30$~K, many rotational levels of the target
molecules are excited. They obey the Boltzmann distribution in typical
experimental conditions. Therefore, this distribution was applied in the
calculations of the formation rates for $T=300$~K. A~significant
difference of the average $dt\mu$ formation rate in the states $K=0$ and
$K=1$ of D$_2$ (see Table~\ref{tab:rate_K}) is due to a~strong increase
of the dominant nuclear formation rate $J=1\to{}J=0$ in the interval
0--0.1~eV and a greater internal kinetic energy of D$_2$ in the state
$K=1$. The contributions from different rotational states~$J$ of the
created muonic molecules to the total nonresonant formation rate
$\Lambda(T)$ are shown in Table~\ref{tab:rate_J} for 30 and 300~K.  Here
the Boltzmann distribution of the rotational levels of the target
molecules is assumed for the both temperatures. One can see that the
molecular rates $\Lambda(T)$ of nonresonant $dd\mu$ formation (in HD and
D$_2$) and $dt\mu$ formation (in D$_2$ and~DT) display a~significant
isotope effect, in particular at the lower temperature.

The formation rates of muonic molecules, which were calculated and
measured in various experiments, are compared in
Table~\ref{tab:th_vs_exp}.
\begin{table*}[htb]
  \begin{center}
    \caption {Experimental and calculated rates [$10^6$~s$^{-1}$] of
      nonresonant formation of the muonic hydrogen molecules in
      different hydrogenic targets.}
    \label{tab:th_vs_exp}
    \newcolumntype{.}{D{.}{.}{1.8}}
    \newcolumntype{,}{D{.}{.}{1.2}}
    \begin{ruledtabular}
      \begin{tabular}{c|.lc|,lr}
        \multicolumn{1}{c|}{Muonic} & \multicolumn{3}{c|}{Experiment} &
        \multicolumn{3}{c}{Theory}\\
        \multicolumn{1}{c|}{molecule} & \multicolumn{1}{c}{Rate} &
        \multicolumn{1}{c}{Conditions} & \multicolumn{1}{c|}{Ref.} &
        \multicolumn{1}{c}{Rate} & \multicolumn{1}{c}{Authors} &
        \multicolumn{1}{c}{Ref.} \\
\hline\noalign{\vskip2pt}
& 0.6^{+0.8}_{-0.5} & Gas, 300~K$^{*}$ &\cite{dzhe62}
& 2.6 & Zel'dovich and Gershtein & \cite{zg} \\
& 1.89\pm 0.20 & Liquid, 22~K$^{*}$ &\cite{bles63}
& 3.9 & Cohen et al. & \cite{cohe60} \\
& 2.55\pm 0.18 & Liquid, 22~K$^{*}$ &\cite{confo64}
& 2.20 & Ponomarev and Faifman & \cite{pono76} \\
 $pp\mu$
& 2.74\pm 0.25 & Gas, 300~K$^{*}$&\cite{budya68}
& 1.80 & Faifman & \cite{mpf89} \\
& 2.34\pm 0.17 & Gas, 300~K$^{*}$ &\cite{byst76}
& 1.81  & Present work, 22~K & \\
& 3.21\pm 0.24 & Solid, 3~K &\cite{mulh96}
& 1.80 & Present work, 300~K &  \\
& 2.01\pm 0.09 & Gas, 300~K &\cite{mucap15} & & & \\
\hline\noalign{\vskip2pt}
& & &
& 1.3 & Zel'dovich and Gershtein&\cite{zg}\\
& 5.8 \pm 0.3 & Liquid, 22~K$^{*}$ & \cite{bles63}
& 3.0 &Cohen et al.&\cite{cohe60} \\
& 6.82\pm 0.25 & Liquid, 22~K$^{*}$ & \cite{confo64}
& 5.91& Ponomarev and Faifman&\cite{pono76} \\
 $pd\mu$
& 5.53\pm 0.16 & Gas, 300~K$^{*}$ & \cite{byst76}
& 5.63& Faifman&\cite{mpf89} \\
& 5.9\pm 0.9 & Liquid, 22~K$^{*}$ & \cite{bertl83}
& 5.63 & Present work, 22~K & \\
& 5.6\pm 0.2 & Liquid, 22~K$^{*}$ & \cite{peti90}
& 5.58 & Present work, 300~K & \\
\hline\noalign{\vskip2pt}
& & & & 0.4 & Zel'dovich and Gershtein&\cite{zg}\\
& & & & 6.49& Ponomarev and Faifman&\cite{pono76}\\
 $pt\mu$
& 7.5\pm 1.3 & Liquid, 23~K$^{*}$ & \cite{baum93}
& 6.38 & Faifman&\cite{mpf89} \\
& & & & 6.38 & Present work, 22~K & \\
& & & & 6.32 & Present work, 300~K & \\
\hline\noalign{\vskip2pt}
& & & & 0.65& Zel'dovich and  Gershtein&\cite{zg}\\
& 1.8 \pm 0.6 & Liquid, 23~K & \cite{breu87}
& 2.96 & Ponomarev and Faifman&\cite{pono76}\\
 $tt\mu$
& 2.3 \pm 0.6 & Solid, 16~K & \cite{matsu99}
& 2.64 & Faifman&\cite{mpf89}  \\
& 2.84\pm 0.32 & Liquid, 22~K & \cite{bogd09}
& 2.69 & Present work, 22~K & \\
& & & & 2.64 & Present work, 300~K & \\
      \end{tabular}
    \end{ruledtabular}
  \end{center}
\end{table*}%
Here the values of the rates from Refs.~\cite{zg,cohe60} and
Refs.~\cite{mpf89,pono76} are quoted at collision energy near zero, and
$\varepsilon=0.04$~eV, respectively. The value of $\lambda_{pp\mu}$ from
the work~\cite{cohe60} has been corrected due to the updated value of
the density of nuclei of liquid hydrogen $N_0=4.25\cdot10^{22} cm^{-3}$,
used in the present work, and in accordance with the remark of
review~\cite{zg}, where is pointed out, that the corrected value should
be two times less. The label ``Present work'' denotes the average rate
for the 22-K or 300-K Maxwell distribution of muonic-atom energies. The
asterisk superscript denotes an assumed value, when the experimental
temperature is not explicitly given in a corresponding reference.  As
can be seen from this table, for the majority of muonic molecules, the
calculated rates are in good agreement with the experimental data.  Only
the value of $pp\mu$ formation rate, which is topical due to the
forthcoming experiments~\cite{adam12,crema2017}, is a special case
because of the disagreements between the theory and experiments.
Moreover, the experimental data measured in different gaseous, liquid
and solid hydrogen targets differ among themselves and are not
sufficiently consistent. Therefore it is rather complicated to make a
conclusion about the degree of agreement between the experimental and
the theoretical values of the $pp\mu$ formation rates.

\section{Conclusions}
\label{sec:concl}

A~role of the internal nuclear motion in the nonresonant formation of
muonic hydrogenic molecules has been considered. In general, this motion
leads to a~significant smearing of the energy-dependent formation rates
$\Lambda(\varepsilon)$ in the $a\mu+BX$ system, when the corresponding
input rates~$\lambda(\varepsilon_b)$ calculated in the $a\mu+b$
system~\cite{mpf89} strongly change within the energy intervals
comparable with the magnitude of vibrational quanta of the target
molecules $BX$ ($\sim{}0.1$~eV). In particular, this effect is important
in the case of nonresonant $dd\mu$ and $dt\mu$ formation at
$\varepsilon\lesssim{}0.1$~eV, which significantly affects the
steady-state kinetics of $\mu$CF processes. Also, an appreciable
isotopic effect in $dd\mu$ nonresonant formation in D$_2$ and HD gas has
been found. Therefore, accurate simulations of various low-energy muonic
processes in hydrogenic molecular targets require the use of the
nonresonant formation rates with the nuclear-motion effect taken into
account.


\begin{acknowledgments}
  The authors are grateful to Prof.\ L.~I.~Men'shikov for helpful
  discussions.
\end{acknowledgments}



\end{document}